*Title*

Computers in Secondary Schools, Educational games

*Affiliation*

Margarida Romero

Laboratoire d'Innovation et Numérique pour l'Education

Université Côte d'Azur, Nice, France

Margarida.Romero@univ-cotedazur.fr

*Synonyms*

Serious games

*Main text*

*Introduction*

This chapter introduces educational games in secondary schools. Educational games include three main types of educational activities with a playful learning intention supported by digital technologies: educational serious games, educational gamification and learning through game creation. Educational serious games are digital games that support learning objectives. Gamification is defined as the use of "game design elements and game thinking in a non-gaming context" (Deterding, Sicart, Nacke, O'Hara, & Dixon, 2011, p. 13) Educational gamification is not developed through a digital game but includes game elements for supporting the learning objectives. Learning through game creation is focused on the process of designing and creating a prototype of a game to support a learning process related to the game creation process or the knowledge mobilized through the game creation process. Four modalities of educational games in secondary education are introduced in this chapter to describe educational games in secondary education: educational repurpose of entertainment games, serious games, gamification and game design.

*Digital games among secondary learners*

Play is renowned as an essential activity for the development of a child from the youngest age; but parents and educators rise important concerns about the negative effects of digital games on secondary learners. Among these concerns, the addictive behavior in relation to massively multiplayer online role-playing game (MMORPG) is among the worst expected outcomes when adults explain their worries about digital games (Gentile, 2009). Some studies has pointed that use of digital games could be a symptom of depression among young adults (Thomée, Härenstam, & Hagberg, 2012). Using games as a way to escape or due to social pressure is also leading to higher levels of addictive uses of games; however, among the factors which do not seem to be linked with the use of games is the parental control (Giles & Price, 2008). While the general conception of digital games is based on entertainment, teenagers and young adults also have expectations about the use of gender affects in the pedagogical expectations on Digital Game Based Learning (DGBL), which differs according to their gender; Karakus, Inal and Cagiltay (2008, p. 2522) observe that "female students expect games to have instructive elements, while males desire elements that are entertaining, competitive, and multi-player". In digital games, male player shows higher skills and time spent in secondary education (Dindar, 2018) and High School (Chou & Tsai, 2007). The representation of age and gender differences in the use of (serious) games leads to stereotypes that are overcome by recent studies on the diversity of playing activity across genders and ages (Loos & Zonneveld, 2016).

*Educational Serious Games (ESG) in secondary schools*

Playing is often perceived as a hobby or an unproductive waste of time (Mackereth & Anderson, 2000). In the last few years, authors, editors and researchers in game-based learning (GBL) have adopted the term serious game to pinpoint the serious nature that a game can have in the development of learning, health or other "serious" objectives considered by adults as profitable. Serious games permit to develop playful learning situations related to disciplinary areas and to 21st century skills such as collaboration, problem solving and creativity (Connolly, Boyle, MacArthur, Hainey, & Boyle, 2012). Among serious games, those having learning objectives could be named under the name of educational serious games (ESG). In secondary education, ESG are developed within the different disciplines of the curriculum such mathematics (Ke, 2013), sciences and technology (Barma & Daniel, 2017; Hodges, Wang, Lee, Cohen, & Jang, 2018), history and language learning (Patino, Proulx, & Romero, 2016). History serious games such *Frequency 1550* (Huizenga, Admiraal, Akkerman, & Dam, 2009) possess a game universe allowing to engage the learners in Amsterdam at 1550. Other popular digital games in secondary schools such *ClassCraft* are based on popular entertainment games (Sanchez, Young, & Jouneau-Sion, 2017).

### *Digital Game-Based Learning in secondary education*

Within the educational games, this section introduces Digital Game-Based Learning (DGBL), serious games and educational gamification in secondary schools including four types of activities*:*

1. *Educational repurpose or 'modding' of entertainment games*. The pedagogical use of digital games that were not specifically conceived for learning or for the pedagogical use of games without educational intentions, for example, the use of *Minecraft* to work on social sciences (citizenship) or mathematical concepts (Hill, 2015; Isiksal & Askar, 2005).

2. *Educational serious games*. The pedagogical use of educational serious games (ESG) conceived with a « serious » educational goal, such the mobile city game called *Frequency 1550*, developed to help secondary education learners to develop historical knowledge of medieval Amsterdam (Huizenga et al., 2009). Within the last years, the use of augmented reality (AR) and virtual reality (VR) in educational serious games in the field of sciences and technology has allowed to approach abstract concepts in a dynamic way through games specifically developed for learning concepts such electromagnetism (Barma & Daniel, 2017) or chemistry (Hodges et al., 2018) in secondary education. Some ESG are based on simulations, defined as "analogies of a real-world situation" (Prensky, 2001, p. 128), in which the learner has the possibility to play with a model or micro-world in order to better understand a system.

3. *Educational gamification* engages game mechanics in an educational context, such in the field of science education (Su & Cheng, 2015). In some contexts, "framing an activity as a game" has induced the learner to a similar engagement and psychological relation to the activity as using game mechanics (Lieberoth, 2015). In terms of learning performance, gamification has showed mixed results (Seaborn & Fels, 2015), specially when the gamification is based in game elements such scores and leaderboards (Toda, Valle, & Isotani, 2017).

4. *Game design and creation*. In this case, the interest resides in the interdisciplinary design procedure of a game. Creating a game is considered as a knowledge modeling activity, which can engage in secondary level learners in same-age or intergenerational game design activities (Cucinelli, Davidson, Romero, & Matheson, 2018).

*Conclusion*

Learning through gaming is not limited to children, and secondary education has been integrating educational serious games (EGS), repurposed entertainment games, gamification systems and game design and creation activities within the last years. While results points to certain benefits in terms of performance in certain tasks, the most important aspect is the learners' engagement and the playful intention which is shared within these different initiatives. Research studies in the next years should continue to study the impact and effects of EGS, repurposed entertainment games, gamification systems and game design and creation activities taking into account not only the specific game components such the Learning Mechanics and Game Mechanics (Arnab et al., 2015; Proulx, Romero, & Arnab, 2016) but also the state-of-the-art frameworks and methodologies within the different disciplines concerned by DGBL studies: from the learning sciences and psychology (Boyle et al., 2016), computer sciences and Human Computer Interaction (Romero, Usart, Popescu, & Boyle, 2012), narrative and storytelling (Boyle et al., 2014) and other disciplinary and cross-disciplinary studies. The development of empirical studies including learning analytics adapted to DGBL leads to the emergence of the playing analytics (Sanchez & Mandran, 2017), which can contribute to better understand the learning process within the game based learning activities developed in digital environments. Through the playing analytics, data could be used to develop adaptative systems embracing the potential of machine learning for education in the next generation of educational games for secondary education but also in other educational levels.

*Cross-References*

*References [max. 30]*